\documentclass[letter]{aa} 
\usepackage[varg]{txfonts}
\usepackage{graphicx}
\usepackage{amsmath}
\usepackage{comment}
\usepackage[stable]{footmisc}
\usepackage{caption}
\usepackage{afterpage}
\usepackage{media9}

\usepackage{xcolor}
\usepackage[colorlinks,breaklinks=true,plainpages=false,citecolor=blue,linkcolor=blue,urlcolor=blue,bookmarksopen=true,bookmarksnumbered=false,bookmarksdepth=5]{hyperref}

\usepackage{enumitem}
\usepackage[normalem]{ulem} 

\newenvironment{tightEq}{%
	\begingroup
	\setlength{\abovedisplayskip}{4pt}%
	\setlength{\belowdisplayskip}{4pt}%
	\setlength{\abovedisplayshortskip}{4pt}%
	\setlength{\belowdisplayshortskip}{4pt}%
}{\endgroup}

\newenvironment{largerEq}{%
	\begingroup
	\setlength{\abovedisplayskip}{6pt}%
	\setlength{\belowdisplayskip}{6pt}%
	\setlength{\abovedisplayshortskip}{8pt}%
	\setlength{\belowdisplayshortskip}{8pt}%
}{\endgroup}

\begin{document}

\title{\emph{Eppur si eclissa}: Eccentric low-mass companions and time-in-dust selection explain long secondary periods}

\author{L. Decin\inst{1} 
    \and
O. Vermeulen\inst{1}
	\and
M. Esseldeurs\inst{1}
	\and
F.\ A.\  Driessen\inst{1}
	\and
C. Landri\inst{1}
 \and
  D. Dionese\inst{2,3}
 \and
 L. Siess\inst{3}
 \and
 D.\ M.\ Skowron\inst{2}
}

\offprints{leen.decin@kuleuven.be}

\institute{
	 Institute of Astronomy, KU Leuven, Celestijnenlaan 200D, 3001 Leuven, Belgium; \url{leen.decin@kuleuven.be}  
	 \and
		 Astronomical Observatory, University of Warsaw, Al.\ Ujazdowskie 4, 00-478 Warsaw, Poland
	 \and
	 	 Institut d'Astronomie et d'Astrophysique, Universit\'e libre de Bruxelles (ULB), CP 226, 1050 Brussels, Belgium 
}

\date{Received October 10, 2025; accepted October 28, 2025}

\abstract{
	\textit{Context.} Long Secondary Periods (LSPs) are observed in \mbox{$\sim$1/3} of pulsating red giants yet remain unexplained. Four key observational constraints anchor the discussion: (i)~a \mbox{$\sim$30\%} occurrence rate in semi-regular variable AGB stars  (SRVs) with a much lower rate (or absence) in regularly pulsating Mira-type AGB stars (Miras), (ii)~\mbox{$\sim$50\%} of LSP stars show a secondary mid-IR minimum; (iii)~Keplerian fits to radial-velocity (RV) curves favour the argument of periastron $\omega>180^{\circ}$; and (iv)~the RV--light curve phase lag clusters around $-\pi/2$. \\
	\textit{Aims.} We test whether a close-in, eccentric low-mass companion that only spends part of its orbit within the giant's dust-formation (wind-launching) zone can match all four empirical facts.\\
	\textit{Methods.} Guided by observed RV amplitudes and periods of $\sim$500\,--\,1500 days, we adopt a companion's mass $M_2\!\in\![0.08,0.25]$\,M$_\odot$, orbital separation $a\!\in\![1.5, 3]$\,au,  eccentricty $e\!\le\!0.6$, and take the dust condensation radius $R_{\rm cond}\!\sim\!2.5$--$3$\,au for SRVs (larger for Miras via scaling with luminosity). We compute the time-in-dust fraction $f_{\rm dust}$ (time with $r\!\ge\!R_{\rm cond}$) and apply line-of-sight criteria: an LSP requires orbital inclination $i\!\ge\!i_{\rm LSP}$ and $f_{\rm dust}\!\ge\!f_{\min}$; a secondary mid-IR minimum interpreted as secondary eclipse further needs $i\!\ge\!i_{\rm ecl}\!>\!i_{\rm LSP}$ and superior conjunction. We test the first three empirical facts analytically, then model the RV–light phase offset with 3D hydrodynamical simulations.\\
	\textit{Results.}  Our proposed scenario  explains the observed excess of $\omega\!>\!180^\circ$. For SRV-like parameters we obtain an LSP detectability of $\sim\!31.6\pm0.1\%$, while Mira-type conditions yield $\sim\!3.0\pm0.1\%$; for both scenarios the conditional secondary mid-IR eclipse fraction is $\sim\!44\%$. 
	Our hydrodynamical models place the optical-depth peak just downstream of the companion near apastron, then shift it to  $\sim$90--225$^\circ$ phase offsets later in the orbit -- consistent with the RV–light offsets. \\
	\textit{Conclusions.} A  time-in-dust geometric selection for low-mass companions in close eccentric orbits explains the four key empirical facts constraining the LSP mechanism. }


\titlerunning{Explaining the statistics behind the LSP phenomenon}

\maketitle

\section{Introduction}\label{Sec:intro}

The long-secondary-period (LSP) phenomenon in red giants was first noted by \cite{OConnell1933}. LSPs occur in \mbox{$\sim$1/3} of pulsating red giants, with periods  $\sim$5$-$10 times the primary pulsation period. In time-domain surveys, LSP stars delineate sequence D in the period–luminosity ($P$-$L$) plane \citep{Wood1999}. The physical origin remains debated, with two leading hypotheses: (i)~non-radial oscillatory convective modes in the outer atmosphere \citep{Saio2015,Takayama2020}, and (ii)~a binary with a close-in, low-mass companion and a \emph{co-orbiting} dusty cloud that periodically obscures the giant \citep{Wood2004,Soszynski2014,Soszynski2021}.

Four empirical facts strongly constrain viable models. First, LSPs are common among semi-regular AGB variables (SRVs, typically double-mode pulsators, with periods between $\sim$60\,--\,250 days) but rarer among the often more luminous, regularly pulsating Mira-type AGB stars (Miras, single-mode pulsators, periods between $\sim$250\,--\,1\,000 days) \citep{Soszynski2022}. Miras typically have larger pulsation-enhanced dust-driven wind velocities and mass-loss rates than SRVs owing to their higher luminosities. 
Second, about half of LSP stars exhibit secondary mid-IR minima without systematic phase lag between the optical and primary mid-IR  minima \citep{Soszynski2022}. 
Third, Keplerian fits to LSP radial-velocity (RV) curves yield a markedly non-uniform distribution of the argument of periastron, $\omega$, clustered at values $>180^\circ$ (median $\sim$227$^\circ$), i.e., at periastron the red giant is closest to the observer and the lower-mass companion is farther away \citep{Wood2004,Nicholls2009}.
Fourth,  observed phase lags between the RV and $I$ band photometric variations cluster near $\Delta\phi\!\approx\! -\pi/2$, implying that a potential companion must be $\sim$180$^\circ$ out of phase with the dust and gas that induce the  brightness minimum of the LSP \citep[][Soszy\'nski et al., in prep.]{Goldberg2024}.

Current oscillatory convective-mode calculations struggle to reproduce sequence~D periods under standard convection parameters and do not naturally account for the combined RV, colour--magnitude, and mid-IR eclipse phenomenology. By contrast, the first two points emerge naturally from  binary--dust geometries, whereas the third and fourth points have been cited as a challenge against the binary picture \citep{Nicholls2009, Goldberg2024}.
More recently, the binary interpretation has been strengthened by the finding that $\sim$50\% of  LSP stars exhibit mid-IR secondary minima \citep{Soszynski2021}, a natural signature of a dusty structure co-orbiting with a companion and producing a secondary eclipse. Meanwhile, RV surveys report full amplitudes of only a few km\,s$^{-1}$  \citep[e.g.,][]{Hinkle2002,Wood2004,Nicholls2009}. Interpreted as orbital motion, this implies companions near the brown-dwarf/very-low-mass main-sequence boundary ($M_2\!\sim\!0.08\text{--}0.25\,\mathrm{M}_\odot$) on AU-scale orbits ($a\!\sim\!1.5\text{--}3$\,au), with typical eccentricities around $e\!\sim\!0.3$ when fitted with Keplerian models \citep{Nicholls2009}; see App.~\ref{Sec:parameters}. This sits in the “brown-dwarf desert”, i.e.\ a relative paucity of companions in this mass range around solar-type stars \citep[e.g.,][]{McCarthy2004,Grether2006}, apparently at odds with the high ($\sim$30\%) LSP incidence among low-mass stars as they ascend the AGB.

One reconciliation proposed by \citet{Soszynski2021} is that many LSP companions did not form at their present mass: planets orbiting AGB stars can accrete from the stellar wind and/or via direct primary-to-companion mass transfer, growing into brown dwarfs or even very low-mass stars \citep{Retter2005}. This view meshes with demographics showing that planets around evolved hosts tend to be more massive than those around main-sequence stars \citep{Jones2014,Niedzielski2015}. Updated occurrence studies also point in this direction: giant planets at a few au are relatively common in field FGK samples \citep{Fulton2021}, and a sample of low-luminosity giants (median mass of 1.21$\pm$0.16\,$M_\odot$) shows  $33.3^{+9.0}_{-7.1}\%$ occurrence rate for Jovian planets within 5\,au \citep{Jones2021}. For higher-mass primaries (initial mass $\ga$1.5\,$M_\odot$) the close-binary fraction around giants will be higher due to the primordial multiplicity-mass relation. This suggests that the fraction of AGB progenitors hosting a low-mass companion capable of producing a dusty wake at $a\!\sim\!1$--few au is substantial, plausibly $f_{\rm comp}\!\sim\!40-60\%$ (or higher).  Writing the SRV LSP incidence as 
\(
f_{\rm LSP}\!=\!f_{\rm comp}\times \big\langle P(\mathrm{detect}\mid\mathrm{comp})\big\rangle,
\)
the observed $f_{\rm LSP}\!\approx\!0.30$ then implies an order-of-magnitude requirement 
\(
\big\langle P(\mathrm{detect}\mid\mathrm{comp})\big\rangle\!\sim\!0.7\text{--}0.9,
\)
which we will use as an empirical benchmark to constrain the geometry in the sections that follow.

The unresolved challenge is that Keplerian binary fits reproduce the observed RV amplitudes but, for random orientations, predict a uniform distribution of \(\omega\) \citep{Nicholls2009}; the observed excess at \(\omega\!>\!180^\circ\)  is therefore puzzling. Moreover, the binary–eclipse picture predicts an RV–\(I\)-band phase lag of \(\Delta\phi\!\approx\!+\pi/2\) at inferior conjunction (companion in front; see Fig.~\ref{Fig:schematic_lag}), contrary to the observed phase lag which  mainly lies between $-\tfrac{1}{5}\pi$ and $-\tfrac{4}{5}\pi$ \citep[see Fig.~3 of][]{Goldberg2024}.
Here we argue that this is a selection effect set by where dust forms around the AGB star and by the companion spending only part of its orbit within that dust-formation region. We propose a simple geometric–selection model in which an eccentric, close-in low-mass companion captures most dust near apastron, producing a trailing spiral wake whose  optical depth then peaks slightly downstream of the companion. As the orbit advances, the dominant optical depth column shifts to the opposite spiral arm, nearly anti-phased with the companion. The resulting time-in-dust gate, coupled with line-of-sight projection, can explain all four key LSP observational constraints.

\vspace*{-1em}
 \section{Model and geometric analysis}
 
 \subsection{Proposed model and the $\mathbf{\omega}$ bias}\label{Sec:model}
 
 \afterpage{\noindent
	\begin{minipage}{\linewidth}
		\centering
		\captionsetup{type=figure,width=\linewidth}
		\vspace*{-1em}
		\includegraphics[width=.9\linewidth]{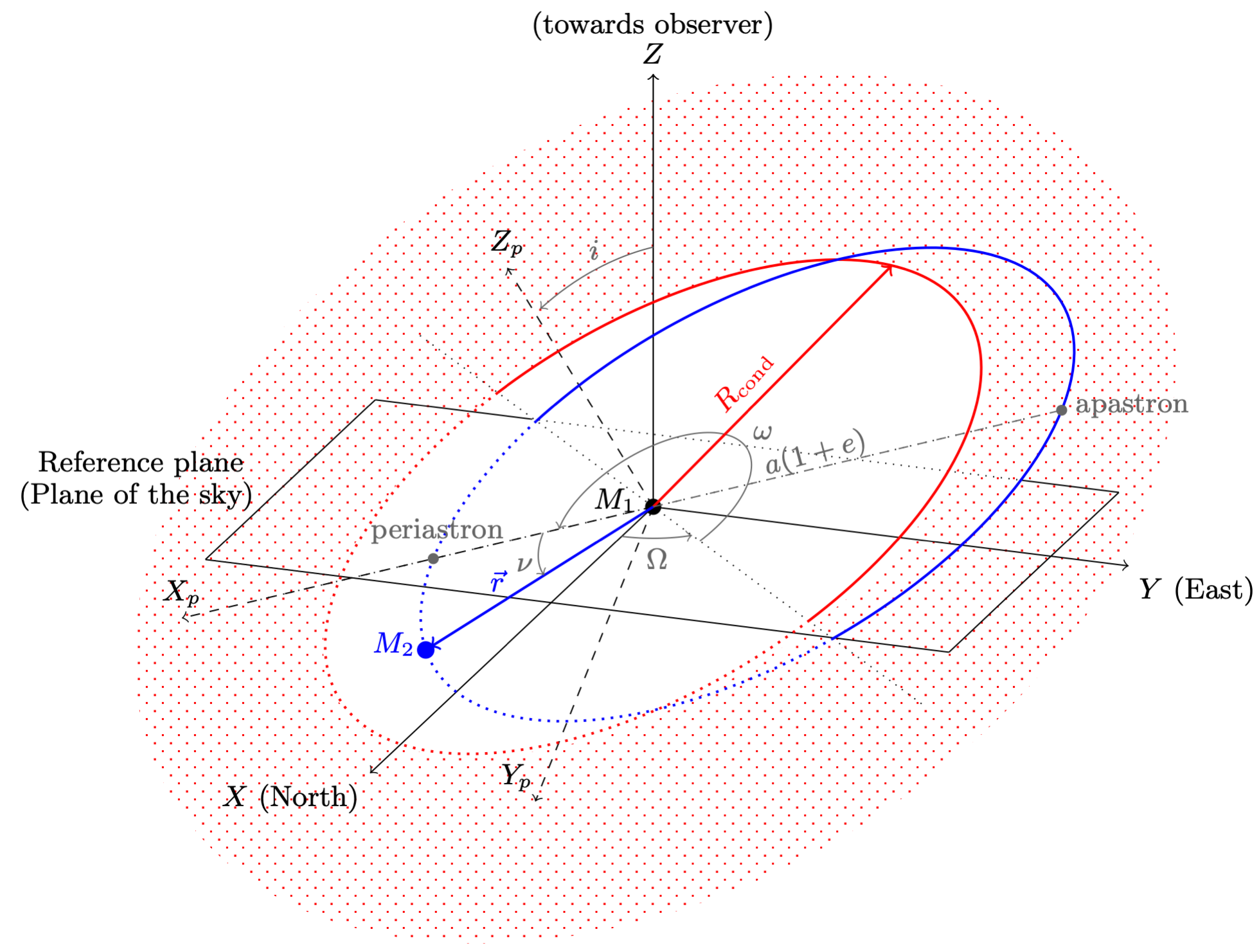}
		\vspace*{-.5em}
		\captionof{figure}{\textbf{Elliptic orbit of a low-mass companion (blue) through the AGB dust-forming region (red).}
			The orbital-plane frame $(X_p,Y_p,Z_p)$ (dashed) and focal frame $(X,Y,Z)$ (solid) are shown. The position vector $\vec r$ from $M_1$ to $M_2$ is set by $\Omega$ (longitude of the ascending node), $i$ (inclination), $\omega$ (argument of periastron; here $225^\circ$)\footnotemark, $a$ (semi-major axis), $e$ (eccentricity), and $T_0$ (time of periastron passage), alongside the true anomaly $\nu$. The companion and its dusty wake cross the spherical dust zone (red dotted) over an apastron-centred phase interval.}
		\label{fig:geom}
	\end{minipage}
\noindent\footnotetext{By convention, RV solutions quote the argument of periastron of the observed star, \(\omega\); the companion’s is \(\omega_{\rm c}=\omega+180^\circ\). For values of $\omega>180^\circ$, the red giant is closest to the observer at periastron.}
}

 In AGB systems, dust condenses at radii of a few stellar radii ($R_\star\!\sim\!1\text{--}1.5$ au), where radiation pressure on grains helps drive the wind \citep{Hofner2018}. Close-in companions lie in the dust-free cavity for $r<R_{\rm cond}$; conversely, only $r\ge R_{\rm cond}$ samples the dust-formation zone.
 In short-period \emph{eccentric} binaries, the periastron may lie inside the dust-free cavity and the companion intersects the dust-formation region only over a limited portion of the orbit, with residence time biased towards apastron  (Fig.~\ref{fig:geom}). For $e=0.3$, the ratio of orbital speeds is $v_{\rm p}/v_{\rm a}=(1+e)/(1-e)\approx 1.86$. Therefore, once $r \ge R_{\rm cond}$ is reached, the longer residence time near apastron -- together with the denser dusty wake -- maximises the line-of-sight optical depth near the apastron sector. Light-curve modulations caused by dust captured in the companion’s gravitational wake will then be stronger. 
 For configurations with $\omega>180^\circ$ (as in Fig.~\ref{fig:geom}), the companion’s passage through the dusty zone occurs at \emph{inferior conjunction}, producing a deeper light-curve minimum than for $\omega<180^\circ$, where inferior conjunction typically finds the companion in the dust-free region with a much weaker spiral wake (see also Sect.~\ref{Sec:hydro}). 
\emph{This scenario naturally explains the observed bias in $\omega$ as an observational selection effect, not an intrinsic property of binary orbital configurations.}

Our scenario assumes eccentric orbits, supported by \citet{Nicholls2009} and the (\(e\!-\!\log P\)) diagram for post-AGB binaries with main-sequence companions, which shows large eccentricities for \(P\!\gtrsim\!100\)~d \citep{VanWinckel2025}. For companions down to \(\sim\!0.1\text{--}0.2\,M_\odot\), the (\(e\!-\!\log P\)) distribution at \(P\!\gtrsim\!100\)~d remains broad -- with a large fraction  at \(e\!\gtrsim\!0.1\) \citep{Moe2017}. Because tidal circularization is far less efficient for such low-mass companions \citep{Zahn1977}, eccentric orbits are expected.

In what follows, we assess the proposed scenario in the light of the other three empirical facts mentioned in Sect.~\ref{Sec:intro}. Because circumstellar dust around an AGB star can exist only at or beyond the condensation radius, the detectability of a companion’s dusty wake is set by (i)~the fraction of the orbit spent at  $r \gtrsim R_{\rm cond}$ (Sect.~\ref{Sec:fdust}) and (ii)~a favourable line-of-sight geometry (Sect.~\ref{Sec:geometry}). Our goal here is not to map the full parameter space, but to test the viability of the proposed scenario under representative AGB-like conditions.  In Sect.~\ref{Sec:analytical}--\ref{Sec:MonteCarlo} we quantify the resulting detection fractions, constrained by the empirical \(\big\langle P(\mathrm{detect}\mid\mathrm{comp})\big\rangle\!\sim\!0.7\text{--}0.9\). The phase offset between RV and light curves is discussed in Sect.~\ref{Sec:hydro}.

\vspace*{-1em}
\subsection{Dust condensation radii and the crossing condition}\label{Sec:fdust}
For AGB stars, interferometry data place the  dust-condensation radius at a few stellar radii \citep[e.g.][]{Wittkowski2007,Sacuto2013,Karovicova2013}; a
simple scaling being
\tightEq
\begin{equation}
	R_{\rm cond} \simeq \left(\frac{L_\star}{16\pi \sigma T_{\rm cond}^4}\right)^{1/2}
	\,\sqrt{\frac{Q_{\rm abs}(T_\star)}{Q_{\rm abs}(T_{\rm sub})}},
\end{equation}
with $T_{\rm cond}$ the dust-condensation temperature ($T_{\rm cond}\!\sim\!1200$--$1500$\,K) and $Q_{\rm abs}$ the dust-grain absorption efficiency \citep{Lamers1999}. Using representative luminosities $L_{\rm SRV}\!\sim\!(3$–$5)\times10^3\,L_\odot$ and $L_{\rm Mira}\!\sim\!10^4\,L_\odot$ gives representative values $R_{\rm cond}\!\sim\!2.5$–$3$\,au for SRVs and $3.5$–$4.5$\,au for Miras.

In our model, \emph{gravitational capture} of circumstellar dust by the companion peaks near apastron (Fig.~\ref{fig:geom}, Sect.~\ref{Sec:hydro}). At the same time, \emph{new-grain production} is maximal in the shock wake driven by the companion near the primary \citep[Fig.~H1 in][]{Danilovich2025}. At other phases the companion resides inside the dust-free cavity, while the previously formed dusty wake is advected outward; the spiral arm broadens, its density contrast fades, and detectability drops. The orbital time fraction spent in the dusty zone is (see App.~\ref{App:time_in_dust})
\begin{equation}
	\label{eq:fdust_closed}
	f_{\rm dust} =
	\begin{cases}
		1, & R_{\rm cond}\le r_{\rm p} = a(1-e),\\[3pt]
		0, & R_{\rm cond}\ge r_{\rm a} = a(1+e),\\[4pt]
		\dfrac{(2\pi-2\theta)+2e\sin\theta}{2\pi}, & \text{otherwise,}
	\end{cases}
\end{equation}
where $\theta=\arccos Y \in [0,\pi]$, and $Y = (1 - R_{\rm cond}/a)/e$.
For SRV-like numbers ($a=2.3$\,au, $e=0.3$, $R_{\rm cond}=2.7$\,au), Eq.~(\ref{eq:fdust_closed}) gives $f_{\rm dust}\approx 0.38$. 
For Mira-like $R_{\rm cond}\!\in\!3.5\text{--}4.5$\,au, one obtains $f_{\rm dust}\!=\!0$; only larger $e$ (and/or larger $a$) yields non-zero dusty phases. \emph{This naturally explains why the LSP phenomenon has a higher occurrence rate in SRVs than in more luminous Miras.}

For Miras, higher mass-loss rates and hence larger circumstellar optical depths reduce the contrast of the companion’s wake against the  optically thick wind, making wake-induced minima harder to isolate. This offers an additional, observation-driven reason  for the lower LSP detection rate in Miras.

\vspace*{-1em}
\subsection{Geometric selection model}\label{Sec:geometry}

 For an LSP to be observable the system must be sufficiently inclined and the companion must spend a minimum fraction of the orbit in the dusty zone:
\begin{equation}
	i \ge i_{\rm LSP}
	\quad\text{and}\quad
	f_{\rm dust} \ge f_{\min}.
	\label{eq:detect_lsp}
\end{equation}
For an average detection fraction \(
\big\langle P(\mathrm{detect}\mid\mathrm{comp})\big\rangle \approx 0.7\text{--}0.9
\), this implies that (see App.~\ref{App_geometry})
\begin{equation}
i_{\rm LSP}\!\approx\!\arccos(0.7-0.9)\!=\!26^\circ\text{--}46^\circ\,.
\label{Eq:i_LSP}
\end{equation} 
  A secondary mid-IR eclipse further requires $i \ge i_{\rm ecl}$ and that the dusty wake then lies  at superior conjunction. Conditioning on systems that already satisfy $i\ge i_{\rm LSP}$, the fraction that also exceed the mid-IR secondary eclipse threshold  is (see App.~\ref{App_geometry})
\begin{equation}
	f_{\rm ecl|LSP}\;\simeq\;\frac{1}{2}\,\frac{\cos i_{\rm ecl}}{\cos i_{\rm LSP}}.
	\label{eq:fecl_lsp}
\end{equation}

\vspace*{-1em}
\subsection{Analytic estimate of the LSP geometric properties}\label{Sec:analytical}

\paragraph{Constraining $i_{\rm LSP}$:}
At conjunction, the sky-projected separation (impact parameter) between the stellar centre and the companion (or its dusty wake) is
\begin{equation}
	b \;\equiv\; r_\perp \;=\; r\,\sqrt{\cos^2 (\omega+\nu) + \sin^2 (\omega+\nu)\,\cos^2 i}\,.
\end{equation}
When conjunction happens near apastron $b\simeq a(1+e)\cos i$. An occultation 
requires the wake to cross the stellar disk,
\begin{equation}
	b \;\le\; R_\star + w_{\rm eff},
\end{equation}
where $R_\star$ is the stellar radius and $w_{\rm eff}$ the effective half-width of the dusty wake.
This yields the inclination threshold
\begin{equation}
	i_{\rm LSP} \;\simeq\; \arccos\left(\frac{R_\star + w_{\rm eff}}{a(1+e)}\right).
\end{equation}

The density scale height of the dusty wake can be approximated by the Bondi–Hoyle–Lyttleton
(BHL) capture radius. 
When the companion’s motion is mostly transverse
to the flow (as near apastron/periastron for a radial wind)
\[
R_{\rm BHL}\;\approx\;\frac{2\,G M_2}{\,v_{\rm orb}^2 + v_w^2 + c_s^2\,}\,.
\]
At apastron, for a chosen representative SRV example  (\(M_1\!=\!1.5\,\mathrm{M_\odot}\), \(M_2\!=\!0.125\,\mathrm{M_\odot}\)
\(a\!=\!2.3\)\,au, \(e\!=\!0.3\),
\(v_w\!=\!10\,\mathrm{km\,s^{-1}}\)) and an adiabatic sound speed $c_s\!=\!2$~km\,s$^{-1}$,
one obtains \(v_{\rm orb}\!=\!v_{\rm a}\!\approx\!18.4~\mathrm{km\,s^{-1}}\) and
\(R_{\rm BHL}\!\approx\!0.5\,\mathrm{au}\). 
However, the BHL approximation is not valid in the case of low wind speeds when the system is in the wind-Roche lobe overflow regime. In that case, the accretion radius can be larger than the Roche Lobe radius of the companion. A convenient upper scale is the
companion’s Hill radius \citep{Decin2020}
\[
w_{\rm eff}\!\approx\!R_H \!\sim\! r\left(\frac{M_2}{3M_1}\right)^{1/3},
\]
which, for our SRV example, 
gives \(R_H \simeq 0.49~\mathrm{au}\) at periastron and \(R_H \simeq 0.91~\mathrm{au}\) at apastron.
In general we adopt a BHL-based scale for the wake and parameterise the effective
half-width as \(
w_{\rm eff}=k\,R_{\rm BHL}\),
with \(k>1\) to reflect that the occulting structure may be broader than the strict accretion cylinder. 
For the chosen representative SRV example, and taking $R_\star\!\approx\!1.5$\,au and $k\!\in\![0.25,2]$ (see  Sect.~\ref{Sec:hydro}),   yields
$i_{\rm LSP} \!\simeq\!33\text{--}57^{\circ}$, \emph{in good agreement overall with the geometric estimate} (Eq.~\ref{Eq:i_LSP}).

\vspace*{-1em}
\paragraph{Secondary mid-IR eclipse and $i_{\rm ecl}$:}
A mid-IR secondary eclipse occurs when, at (near) superior conjunction, the warm IR--emitting core of the dusty wake is partly occulted by the stellar disk. This requires a slightly smaller impact parameter than for detectable optical obscuration because the mid-IR dip is an occultation of the more compact, warm dust emission. We parametrise the effective half-width of the IR–emitting core as $w_{\rm IR}=\zeta\,w_{\rm eff}$. Because the mid-IR emission is centrally concentrated around the warmest part of the wake, $\zeta\lesssim 1$ is expected. The corresponding inclination threshold is
\begin{equation}
	i_{\rm ecl}\;\simeq\;\arccos\left(\frac{R_\star+w_{\rm IR}}{a(1+e)}\right)
	\,=\,\arccos\left(\frac{R_\star+\zeta\,w_{\rm eff}}{r_{\rm a}}\right).
\end{equation}
 Using the same SRV values as before for $k=1$  (hence $i_{\rm LSP}\!\sim\!48^\circ$) and $\zeta\simeq0.6\text{--}0.8$ yields $i_{\rm ecl}\!\approx\!51\text{--}53^\circ$, hence $i_{\rm ecl}$ is typically only a few degrees larger than $i_{\rm LSP}$. Eq.~(\ref{eq:fecl_lsp}) then gives $f_{\rm ecl|LSP}\!\approx\!45\text{--}47\%$, \emph{matching the observed $\sim$50\% of mid-IR eclipse events}.

\vspace*{-1em}
\subsection{Monte-Carlo analysis}\label{Sec:MonteCarlo}
We perform a Monte-Carlo experiment drawing $10^5$ systems per run and repeating the experiment over 125 independent random seeds. For each draw we sample orbital/stellar/wind parameters from the ranges listed below, assign random viewing geometry, and apply the detectability criteria of Sects.~\ref{Sec:fdust}--\ref{Sec:geometry}. 

As will be discussed in Sect.~\ref{Sec:hydro}, $k$ can range between 0.1 and 2, with larger values  further downstream the spiral flow. Sampling $a\!\in\![1.5,3]$~au, $e\!\lesssim\!0.6$ \citep[median $\sim$0.3,][]{Nicholls2009}, 
$M_1\!\in\![1.0,1.5]\,M_\odot$, 
$M_2\!\in\![0.08,0.25]\,M_\odot$, 
$v_w\!\in\![8,12]$~km\,s$^{-1}$, $R_\star\!\in\![1.0,1.5]$~au, $R_{\rm cond}\!\in\![2.5,3]$~au, 
$k\!\in\![0.25,2]$, $\zeta\!\in\![0.5,0.9]$, and  $f_{\rm dust}\!\ge\!0.1$,
we obtain an SRV LSP detectability fraction $f_{\rm LSP}\!\approx\!31.6\pm0.1\%$ and a conditional
secondary mid-IR eclipse fraction $f_{\rm ecl|LSP}\!\approx\!44.8\pm0.3\%$.
 Increasing $a\!\in\![3.5,4.5]$~au and $v_w\!\in\![10,20]$~km\,s$^{-1}$ to mimic Mira-type conditions yields $f_{\rm LSP}\!\approx\!3.0\pm0.1\%$ and a conditional
 secondary mid-IR eclipse fraction $f_{\rm ecl|LSP}\!\approx\!44.0\pm0.9\%$.
The eclipse fraction near one-half emerges naturally from the geometry, while the overall LSP
incidence decreases as $R_{\rm cond}/a$ increases (Mira-like conditions), consistent with the
observed SRV--Mira contrast.

\vspace*{-1em}
\subsection{Phase offset between RV and light curves}\label{Sec:hydro}

\begin{figure}[htp]
	\centering
	\vspace*{-1.3em}
	\includegraphics[width=.83\linewidth]{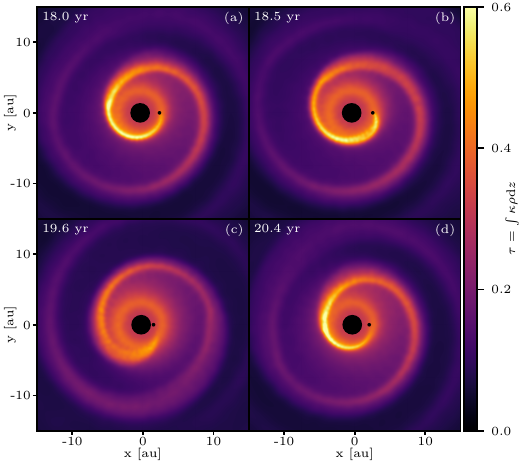}
	\vspace*{-.5em}
	\caption{Optical-depth maps along the $z$ axis for a slice through the orbital plane for a binary setup with $M_1\!=\!1.5\,M_\odot$, $M_2\!=\!0.125\,M_\odot$, $a\!=\!2.3$~au, $e\!=\!0.3$, $R_{\rm cond}\!=\!3$\,au and mass-loss rate of $5\!\times\!10^{-7}$~$M_\odot$\,yr$^{-1}$.  Plots are in the co-moving frame, with the center of mass at  \((0,0, 0)\) and the low-mass companion to the right. Video   available at Video~\ref{Video:optical_depth}.}	
	\label{fig:tau_map}
\end{figure}

\vspace*{-.5em}
To study the origin of the observed phase offset between  RV curves and  \(I\)-band photometry, we run three-dimensional smoothed-particle hydrodynamics  simulations with \textsc{Phantom} \citep{Price2018}, with numerical binary setup as described in \citet{Malfait2024}. The opacity is calculated using the analytical expression proposed by \citet{Bowen1988} with the dust equilibrium temperature being calculated under the Lucy approximation, as implemented by \citet{Esseldeurs2023}. Four representative  snapshots are shown in Fig.~\ref{fig:tau_map}. In this simulation, the effective half-width of the inner spiral wake  is $w_{\rm eff}\!\simeq\!0.25\text{--}2.0\,R_{\rm BHL}$ depending on the cut angle (see App.~\ref{Sec:width})


As the companion approaches apastron within the dust-formation region, the wake densifies and the optical depth $\tau$ increases; near periastron the wake is more rarefied and the overall \(\tau\) is lower (see  Fig.~\ref{fig:tau_map}, panel~(c)). This  supports our interpretation of the observed \(\omega\!>\!180^\circ\) bias. The locus of maximum optical depth varies with orbital phase. Near apastron, the brightest ridge usually lies in the wake, slightly downstream of the companion. For much of the orbit, however, the dominant peak occurs \(\sim\!90^\circ\)–\(225^\circ\) out of phase with the companion, i.e, the phase offset between the RV and photometric variations occurs at $[-\pi,-\tfrac{1}{4}\pi]$. Because the observed flux integrates extinction along the entire line of sight, this geometry naturally produces \emph{photometric minima that are approximately anti-phased with the companion, consistent with the observed RV–\(I\) phase lag reported by} \citet{Goldberg2024}. 

\vspace*{-1em}
\section{Conclusions}

\vspace{-.5em}
A  geometric and time-in-dust criterion -- set by the orbit's fraction of an eccentric, close-in low-mass companion with $r\!\ge\!R_{\rm cond}$  and by  line-of-sight geometry from a gravitationally focused dusty wake  -- simultaneously accounts for (i)~the $\sim$30\% LSP occurrence rate in SRVs versus the much lower  rate in Miras, (ii)~the $\sim$50\% fraction of LSP stars that show a secondary mid-IR eclipse,  (iii)~the observed excess of fitted  $\omega>180^\circ$, and (iv)~the phase lag between RV and $I$ band light curves clustering around $-\pi/2$. The framework yields testable trends with wind speed, companion mass, and wake breadth.

Our selection model is intentionally minimal and does not account for (i)~time variability of $R_{\rm cond}$, which introduces scatter in the eclipse fraction but does not change the qualitative SRV--Mira contrast or the $\omega$ bias. Moreover, (ii)~the effective occulting width, $w_{\rm eff}$, encapsulates bow-shock compression and the dusty sheath, and is uncertain. The parameters $k$ and $\zeta$ depend on the specific orbital configuration as well as on the abundance and size distribution of the newly formed dust grains. Lowering $k$ reduces the SRV LSP detectability fraction, whereas decreasing $\zeta$ reduces the conditional secondary mid-IR eclipse fraction.
In addition, (iii)~non-radial/convective pulsation modes \citep{Saio2015} may contribute to colour/phase relations in some objects. Future work should couple 3D hydrodynamics and radiative transfer to predict multi-band light curves and RV line-shape diagnostics in function of  the system parameters.

\vspace{-.3em}
\begin{acknowledgements}
L.D., M.E., F.D.\ and C.L.\ acknowledge support from the KU Leuven C1 excellence grant BRAVE C16/23/009, KU Leuven Methusalem grant SOUL METH/24/012, and the FWO research grants G099720N and G0B3823N. L.D.\ acknowledges support from the FWO sabbatical grant and O.V.\ from the FWO PhD grant 1173025N. D.M.S.\ and D.D.\ acknowledge support from the European Union (ERC, LSP-MIST, 101040160). Views and opinions expressed are however those of the authors only and do not necessarily reflect those of the European Union or the European Research Council. Neither the European Union nor the granting authority can be held responsible for them.
\end{acknowledgements}

\largerEq
\vspace*{-2.2em}
\bibliographystyle{aa}
\bibliography{lsp_letter}

\begin{thebibliography}{33}
\expandafter\ifx\csname natexlab\endcsname\relax\def\natexlab#1{#1}\fi

\bibitem[{{Bowen}(1988)}]{Bowen1988}
{Bowen}, G.~H. 1988, \apj, 329, 299

\bibitem[{{Danilovich} {et~al.}(2025){Danilovich}, {Samaratunge}, {Mori},
  {Richards}, {Baudry}, {Etoka}, {Montarg{\`e}s}, {Kervella}, {McDonald},
  {Gottlieb}, {Wallace}, {Price}, {Decin}, {Bolte}, {Ceulemans}, {De Ceuster},
  {de Koter}, {Dionese}, {El Mellah}, {Esseldeurs}, {Gray}, {Herpin}, {Khouri},
  {Lagadec}, {Landri}, {Marinho}, {Menten}, {Millar}, {M{\"u}ller},
  {Pimpanuwat}, {Plane}, {Sahai}, {Siess}, {Van de Sande}, {Vermeulen}, {Wong},
  {Yates}, \& {Zijlstra}}]{Danilovich2025}
{Danilovich}, T., {Samaratunge}, N., {Mori}, Y., {et~al.} 2025, \aap, in press

\bibitem[{{Decin} {et~al.}(2020){Decin}, {Montarg{\`e}s}, {Richards},
  {Gottlieb}, {Homan}, {McDonald}, {El Mellah}, {Danilovich}, {Wallstr{\"o}m},
  {Zijlstra}, {Baudry}, {Bolte}, {Cannon}, {De Beck}, {De Ceuster}, {de Koter},
  {De Ridder}, {Etoka}, {Gobrecht}, {Gray}, {Herpin}, {Jeste}, {Lagadec},
  {Kervella}, {Khouri}, {Menten}, {Millar}, {M{\"u}ller}, {Plane}, {Sahai},
  {Sana}, {Van de Sande}, {Waters}, {Wong}, \& {Yates}}]{Decin2020}
{Decin}, L., {Montarg{\`e}s}, M., {Richards}, A.~M.~S., {et~al.} 2020, Science,
  369, 1497

\bibitem[{{El Mellah} {et~al.}(2020){El Mellah}, {Bolte}, {Decin}, {Homan}, \&
  {Keppens}}]{ElMellah2020}
{El Mellah}, I., {Bolte}, J., {Decin}, L., {Homan}, W., \& {Keppens}, R. 2020,
  \aap, 637, A91

\bibitem[{{Esseldeurs} {et~al.}(2023){Esseldeurs}, {Siess}, {De Ceuster},
  {Homan}, {Malfait}, {Maes}, {Konings}, {Ceulemans}, \&
  {Decin}}]{Esseldeurs2023}
{Esseldeurs}, M., {Siess}, L., {De Ceuster}, F., {et~al.} 2023, \aap, 674, A122

\bibitem[{{Fulton} {et~al.}(2021){Fulton}, {Rosenthal}, {Hirsch}, {Isaacson},
  {Howard}, {Dedrick}, {Sherstyuk}, {Blunt}, {Petigura}, {Knutson}, {Behmard},
  {Chontos}, {Crepp}, {Crossfield}, {Dalba}, {Fischer}, {Henry}, {Kane},
  {Kosiarek}, {Marcy}, {Rubenzahl}, {Weiss}, \& {Wright}}]{Fulton2021}
{Fulton}, B.~J., {Rosenthal}, L.~J., {Hirsch}, L.~A., {et~al.} 2021, \apjs,
  255, 14

\bibitem[{{Goldberg} {et~al.}(2024){Goldberg}, {Joyce}, \&
  {Moln{\'a}r}}]{Goldberg2024}
{Goldberg}, J.~A., {Joyce}, M., \& {Moln{\'a}r}, L. 2024, \apj, 977, 35

\bibitem[{{Grether} \& {Lineweaver}(2006)}]{Grether2006}
{Grether}, D. \& {Lineweaver}, C.~H. 2006, \apj, 640, 1051

\bibitem[{Hinkle {et~al.}(2002)Hinkle, Lebzelter, Joyce, \& Fekel}]{Hinkle2002}
Hinkle, K.~H., Lebzelter, T., Joyce, R.~R., \& Fekel, F.~C. 2002, \aj, 123,
  1002

\bibitem[{{H{\"o}fner} \& {Olofsson}(2018)}]{Hofner2018}
{H{\"o}fner}, S. \& {Olofsson}, H. 2018, \aapr, 26, 1

\bibitem[{{Jones} {et~al.}(2014){Jones}, {Jenkins}, {Bluhm}, {Rojo}, \&
  {Melo}}]{Jones2014}
{Jones}, M.~I., {Jenkins}, J.~S., {Bluhm}, P., {Rojo}, P., \& {Melo}, C.~H.~F.
  2014, \aap, 566, A113

\bibitem[{{Jones} {et~al.}(2021){Jones}, {Wittenmyer}, {Aguilera-G{\'o}mez},
  {Soto}, {Torres}, {Trifonov}, {Jenkins}, {Zapata}, {Sarkis}, {Zakhozhay},
  {Brahm}, {Ram{\'\i}rez}, {Santana}, {Vines}, {D{\'\i}az},
  {Vu{\v{c}}kovi{\'c}}, \& {Pantoja}}]{Jones2021}
{Jones}, M.~I., {Wittenmyer}, R., {Aguilera-G{\'o}mez}, C., {et~al.} 2021,
  \aap, 646, A131

\bibitem[{{Karovicova} {et~al.}(2013){Karovicova}, {Wittkowski}, {Ohnaka},
  {Boboltz}, {Fossat}, \& {Scholz}}]{Karovicova2013}
{Karovicova}, I., {Wittkowski}, M., {Ohnaka}, K., {et~al.} 2013, \aap, 560, A75

\bibitem[{{Lamers} \& {Cassinelli}(1999)}]{Lamers1999}
{Lamers}, H. J.~G.~L.~M. \& {Cassinelli}, J.~P. 1999, {Introduction to Stellar
  Winds} (Cambridge University Press)

\bibitem[{{Malfait} {et~al.}(2024){Malfait}, {Siess}, {Esseldeurs}, {De
  Ceuster}, {Wallstr{\"o}m}, {de Koter}, \& {Decin}}]{Malfait2024}
{Malfait}, J., {Siess}, L., {Esseldeurs}, M., {et~al.} 2024, \aap, 691, A84

\bibitem[{{McCarthy} \& {Zuckerman}(2004)}]{McCarthy2004}
{McCarthy}, C. \& {Zuckerman}, B. 2004, \aj, 127, 2871

\bibitem[{{Moe} \& {Di Stefano}(2017)}]{Moe2017}
{Moe}, M. \& {Di Stefano}, R. 2017, \apjs, 230, 15

\bibitem[{Nicholls {et~al.}(2009)Nicholls, Wood, Cioni, \&
  Soszy{\'n}ski}]{Nicholls2009}
Nicholls, C.~P., Wood, P.~R., Cioni, M.-R.~L., \& Soszy{\'n}ski, I. 2009,
  \mnras, 399, 2063

\bibitem[{{Niedzielski} {et~al.}(2015){Niedzielski}, {Wolszczan}, {Nowak},
  {Adam{\'o}w}, {Kowalik}, {Maciejewski}, {Deka-Szymankiewicz}, \&
  {Adamczyk}}]{Niedzielski2015}
{Niedzielski}, A., {Wolszczan}, A., {Nowak}, G., {et~al.} 2015, \apj, 803, 1

\bibitem[{{O'Connell}(1933)}]{OConnell1933}
{O'Connell}, D.~J.~K. 1933, Harvard College Observatory Bulletin, 893, 19

\bibitem[{{Price} {et~al.}(2018){Price}, {Wurster}, {Tricco}, {Nixon},
  {Toupin}, {Pettitt}, {Chan}, {Mentiplay}, {Laibe}, {Glover}, {Dobbs},
  {Nealon}, {Liptai}, {Worpel}, {Bonnerot}, {Dipierro}, {Ballabio}, {Ragusa},
  {Federrath}, {Iaconi}, {Reichardt}, {Forgan}, {Hutchison}, {Constantino},
  {Ayliffe}, {Hirsh}, \& {Lodato}}]{Price2018}
{Price}, D.~J., {Wurster}, J., {Tricco}, T.~S., {et~al.} 2018, \pasa, 35, e031

\bibitem[{Retter(2005)}]{Retter2005}
Retter, A. 2005, in AAS Meeting Abstracts, Vol. 207, 191.06

\bibitem[{Sacuto {et~al.}(2013)Sacuto, Ramstedt, H{\"o}fner, Aringer, Maercker,
  Olofsson, Wittkowski, Khouri, \& Kerschbaum}]{Sacuto2013}
Sacuto, S., Ramstedt, S., H{\"o}fner, S., {et~al.} 2013, \aa, 551, A72

\bibitem[{Saio {et~al.}(2015)Saio, Wood, Takayama, \& Ita}]{Saio2015}
Saio, H., Wood, P.~R., Takayama, M., \& Ita, Y. 2015, \mnras, 452, 3863

\bibitem[{{Soszy{\'n}ski}(2022)}]{Soszynski2022}
{Soszy{\'n}ski}, I. 2022, in XL Polish Astronomical Society Meeting, ed.
  E.~{Szuszkiewicz} \& A.~{Majczyna}, Vol.~12, 154--157

\bibitem[{Soszy{\'n}ski {et~al.}(2021)Soszy{\'n}ski, Olechowska, Ratajczak,
  Iwanek, Skowron, Mr{\'o}z, Pietrukowicz, Udalski, Szyma{\'n}ski, Skowron,
  Gromadzki, Poleski, Koz{\l}owski, Wrona, Ulaczyk, \& Rybicki}]{Soszynski2021}
Soszy{\'n}ski, I., Olechowska, A., Ratajczak, M., {et~al.} 2021, \apjl, 911,
  L22

\bibitem[{{Soszy{\'n}ski} \& {Udalski}(2014)}]{Soszynski2014}
{Soszy{\'n}ski}, I. \& {Udalski}, A. 2014, \apj, 788, 13

\bibitem[{{Takayama} \& {Ita}(2020)}]{Takayama2020}
{Takayama}, M. \& {Ita}, Y. 2020, \mnras, 492, 1348

\bibitem[{{Van Winckel}(2025)}]{VanWinckel2025}
{Van Winckel}, H. 2025, Galaxies, 13, 68

\bibitem[{Wittkowski {et~al.}(2007)Wittkowski, Boboltz, Driebe, Humphreys,
  Leinert, \& Scholz}]{Wittkowski2007}
Wittkowski, M., Boboltz, D.~A., Driebe, T., {et~al.} 2007, \aa, 470, 191

\bibitem[{{Wood} {et~al.}(1999){Wood}, {Alcock}, {Allsman}, {Alves}, {Axelrod},
  {Becker}, {Bennett}, {Cook}, {Drake}, {Freeman}, {Griest}, {King}, {Lehner},
  {Marshall}, {Minniti}, {Peterson}, {Pratt}, {Quinn}, {Stubbs}, {Sutherland},
  {Tomaney}, {Vandehei}, \& {Welch}}]{Wood1999}
{Wood}, P.~R., {Alcock}, C., {Allsman}, R.~A., {et~al.} 1999, in IAU Symposium,
  Vol. 191, Asymptotic Giant Branch Stars, ed. T.~{Le Bertre}, A.~{Lebre}, \&
  C.~{Waelkens}, 151

\bibitem[{Wood {et~al.}(2004)Wood, Olivier, \& Kawaler}]{Wood2004}
Wood, P.~R., Olivier, E.~A., \& Kawaler, S.~D. 2004, \apj, 604, 800

\bibitem[{{Zahn}(1977)}]{Zahn1977}
{Zahn}, J.~P. 1977, \aap, 57, 383

\end{thebibliography}

\begin{appendix}
	\setcounter{figure}{0}

	\section{Orbital parameters} \label{Sec:parameters}
	
	Published LSP RV curves typically have semi-amplitudes \(K_1\!\sim\!1\text{--}3~\mathrm{km\,s^{-1}}\) (full amplitudes \(\sim 2\text{--}6~\mathrm{km\,s^{-1}}\)) \citep{Hinkle2002,Wood2004,Nicholls2009}. For a representative full amplitude of \(3.5~\mathrm{km\,s^{-1}}\) (i.e. \(K_1=1.75~\mathrm{km\,s^{-1}}\)) and periods \(P\!\sim\!500\text{--}1500~\mathrm{d}\), the spectroscopic mass function
			\begin{equation}
				f(M) \equiv \frac{(M_2 \sin i)^3}{(M_1+M_2)^2}
				= 1.036\times10^{-7}\, K_1^3\, P_{\rm d}\, (1-e^2)^{3/2}\; M_\odot,
			\end{equation}
			(with \(K_1\) in \(\mathrm{km\,s^{-1}}\) and \(P_{\rm d}\) in days) yields
			\(f(M)\!\sim\!(2.4\text{--}7.2)\times10^{-4}\,M_\odot\) for \(e\!\sim\!0.3\) (the median eccentricity reported by \citealt{Nicholls2009}). Assuming \(M_1 \sim 1\text{--}1.5\,M_\odot\) and \(\sin i \sim 1\) gives \(M_2\!\approx\!0.08\text{--}0.14\,M_\odot\) (edge-on minima), rising to \(\sim\!0.12\text{--}0.25\,M_\odot\) for a more typical inclination \(i\!\sim\!60^\circ\). Kepler’s third law then implies \(a\!\sim\!1.5\text{--}3~\mathrm{au}\). In our analysis, we adopt \(e\!\lesssim\!0.6\) as a prior, as derived from the LSP RV fits \citep{Nicholls2009}.

	\section{Time-in-dust gate}\label{App:time_in_dust}
	For an ellipse with semi-major axis $a$ and eccentricity $e$, the orbital radius as a function of eccentric anomaly $E$ is
	\begin{equation}
		r(E)=a\,[1-e\cos E],\qquad M=E-e\sin E,
	\end{equation}
	with $M$ the mean anomaly. Because $M$ advances uniformly in time, we define the time fraction spent in the dusty zone as
	\begin{equation}
		f_{\rm dust}(a,e;R_{\rm cond}) \equiv \frac{\Delta t(r\ge R_{\rm cond})}{P}
		= \frac{\Delta M}{2\pi},
		\label{eq:fdust_def}
	\end{equation}
	where the boundary $r=R_{\rm cond}$ is reached when $1-e\cos E=R_{\rm cond}/a$, i.e.
	\begin{equation}
		\cos E \equiv Y = \frac{1 - R_{\rm cond}/a}{e}.
	\end{equation}
	Hence,
	\begin{equation}
		f_{\rm dust} =
		\begin{cases}
			1, & R_{\rm cond}\le r_{\rm p} = a(1-e),\\[3pt]
			0, & R_{\rm cond}\ge r_{\rm a} = a(1+e),\\[4pt]
			\dfrac{(2\pi-2\theta)+2e\sin\theta}{2\pi}, & \text{otherwise,}
		\end{cases}
	\end{equation}
	where $\theta=\arccos Y \in [0,\pi]$. 
	
	\section{Geometric selection model}\label{App_geometry}
	Photometric and radial-velocity observables (e.g., light-curve depth, mid-IR eclipse visibility, and RV semi-amplitude $K$) are invariant under $(i,\Omega)\mapsto(\pi-i,\Omega+\pi)$. We may therefore fold the inclination to $i\in[0,\pi/2]$ without loss of generality. For an isotropic distribution of orbital planes, $\cos i \sim \mathcal{U}(0,1)$ on $[0,1]$, which implies a probability density function 
		\(
		p(i)=\sin i,
		\)
		and thus the probability of having $i\!\ge\!i_0$
		\[
		\Pr(i\ge i_{\rm LSP})=\int_{i_{\rm LSP}}^{\pi/2}\sin i\,\mathrm{d}i=\cos i_{\rm LSP}
		\]
		If we require an average detection fraction
		\(
		\big\langle P(\mathrm{detect}\mid\mathrm{comp})\big\rangle \approx 0.7\text{--}0.9,
		\)
		then
		\begin{equation}
			i_{\rm LSP} \!\approx\!\arccos(0.7\text{--}0.9)\!=\!26^\circ\text{--}46^\circ.
		\end{equation}
		
\begin{figure*}[htp]
	\centering
	\includegraphics[width=\linewidth]{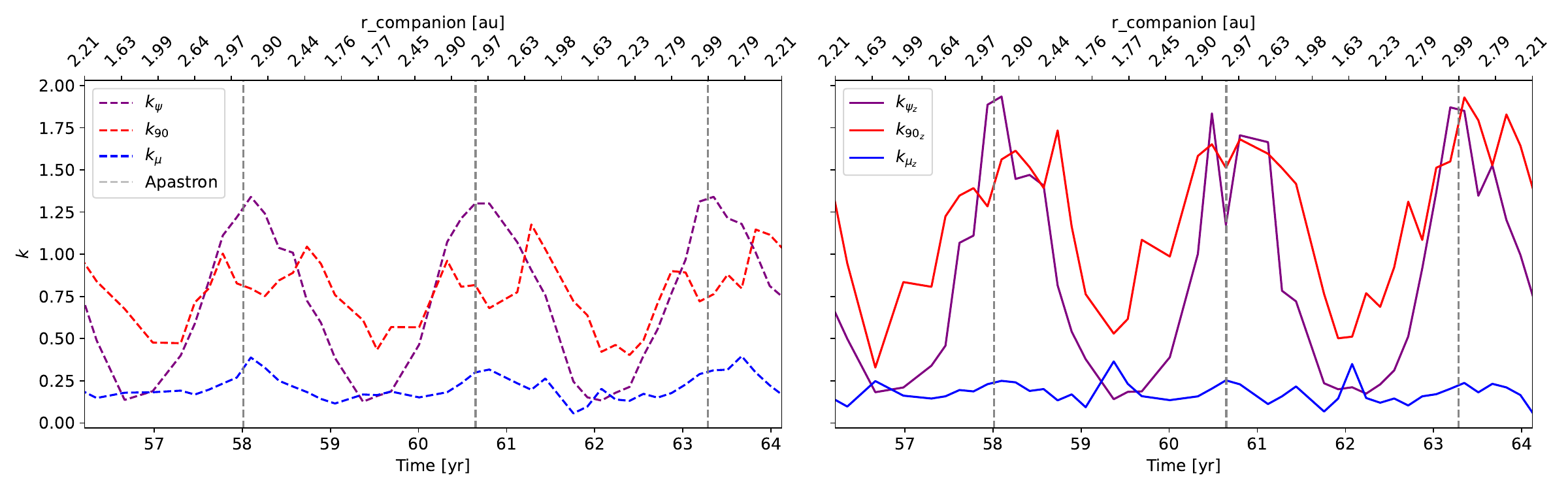}
	\caption{Evolution of \(k \equiv w_{\rm eff}/R_{\rm BHL}\) over \(\sim3\) orbits for cut angles: \(\theta=-\mu\) (red), \(\theta=-\psi\) (purple), and \(\theta=-90^\circ\) (blue). Bottom x-axis: time; top x-axis: orbital radius \(r\). Left panel: rays in the orbital plane. Right panel: rays perpendicular to the plane (\(z\)-direction). In each panel, the apastron passage is indicated with a grey dashed vertical line. Curves are computed from discrete hydro snapshots (only every few time steps retained), so the small-scale jaggedness reflects the sampling cadence rather than intrinsic variability.}
	\label{Fig:k}
\end{figure*}

		Conditioning on systems that already satisfy $i\ge i_{\rm LSP}$, the fraction that also exceed the mid-IR eclipse threshold $i_{\rm ecl}$ (with $i_{\rm ecl}\ge i_{\rm LSP}$; see Sect.~\ref{Sec:geometry}) is
		\[
		\Pr(i\ge i_{\rm ecl}\mid i\ge i_{\rm LSP})
		=\frac{\cos i_{\rm ecl}}{\cos i_{\rm LSP}}.
		\]
		Because a mid-IR secondary dip additionally requires the far-side (superior-conjunction) configuration and, for random nodes, that occurs half the time, we include a factor $\simeq\tfrac{1}{2}$:
		\begin{equation}
			f_{\rm ecl|LSP}\;\simeq\;\frac{1}{2}\,\frac{\cos i_{\rm ecl}}{\cos i_{\rm LSP}}.
			\label{eq:fecl_lsp_app}
		\end{equation}

\section{Effective half-width of the inner spiral arm}
\label{Sec:width}

We measure the effective half-width of the inner spiral wake, $w_{\rm eff}$, from the
hydrodynamical simulation (Sect.~\ref{Sec:hydro}). We define $w_{\rm eff}$ as the density
HWHM: the distance from the ridge (peak density) to where $\rho=\tfrac{1}{2}\rho_{\max}$,
using cuts perpendicular to the local arm tangent. As in Sect.~\ref{Sec:analytical} we write
\[
w_{\rm eff}\simeq k\,R_{\rm BHL}.
\]
In general,
\[
R_{\rm BHL}=\frac{2GM_2}{v_{\rm rel}^2+c_s^2},\qquad
\mathbf v_{\rm rel}\equiv \mathbf v_{\rm wind}-\mathbf v_{\rm comp},
\]
evaluated in the instantaneous rest frame of the companion. For a radial wind and Keplerian
motion,
\[
v_{\rm rel}^2=(v_w-v_r)^2+v_t^2,\qquad
v_r=\frac{\mu_\star}{h}\,e\sin\nu,\quad
v_t=\frac{\mu_\star}{h}\,(1+e\cos\nu),
\]
with
\[
h=\sqrt{\mu_\star a(1-e^2)},\qquad \mu_\star\equiv G(M_1+M_2),
\]
and $G$ the gravitational constant. At the apsides ($\nu=0,\pi$), $v_r=0$ so $v_{\rm rel}^2=v_w^2+v_{\rm orb}^2$.

\smallskip

The inner-arm compression ridge peaks only a few degrees downstream of the radial.
We therefore set the downstream sampling angle by the flow geometry and define
\[
\mu \equiv \arcsin\left(\frac{c_s}{v_{\rm rel}}\right),
\]
so that, with angles measured in the orbital plane from the primary to the companion
(increasing prograde), the downstream limb for a prograde orbit is at
\[
\theta = -\,\mu.
\]
We measure $w_{\rm eff}$ at $\theta=-\mu$ on a cut perpendicular to the arm.

For comparison, we also measure $w_{\rm eff}$ along the arm–normal set by the local
pitch angle $\psi$, being the angle between the arm tangent  and the
azimuthal \citep[see Fig.~6 of][]{ElMellah2020}, with
\[
\tan\psi=\frac{1}{r}\frac{{\rm d}r}{{\rm d}\phi}\;\simeq\;\frac{v_w}{v_{\rm orb}}.
\]
With angles measured from the radial  and increasing prograde, the
arm–normal lies at $\theta=-\psi$. Sampling at $\theta=-\psi$ gives the intrinsic
cross–arm width and, in our setup, falls a few–tens of degrees downstream. Larger $\psi$
corresponds to a more open, radially expanding segment (material launched when the
companion’s angular speed was lower, near apastron), while smaller $\psi$ indicates a
tighter, more wound segment (launched near periastron). We also sample the
lateral trailing direction at $\theta=-90^\circ$ to gauge how the width changes away from
the downstream focusing.

Fig.~\ref{Fig:k} shows that \(k\) is largest near apastron and decreases towards periastron. Over one orbit we find \(k\) to vary between $\sim$0.25 and $\sim$2.0 for $\theta\!=\!-\psi$ and  $\theta\!=\!-90^\circ$, being only between $\sim$0.15 and $\sim$0.25 for $\theta\!=\!-\mu$. 

We note that $R_{\rm BHL}$ is a capture/focusing scale for rectilinear upstream flow, not a lower bound on the morphological thickness of the post-shock ridge. Along the strongly compressed downstream limb ($\theta\!=\!-\mu$) the inner arm behaves as a thin, shock-bounded sheet whose thickness is set by compression and cooling in supersonic flow rather than by $R_{\rm BHL}$. Orbital curvature and shear further concentrate material, while the instantaneous $v_{\rm rel}$ -- largest near periastron -- both reduces $R_{\rm BHL}$ and enhances compression. Consequently, $k\!<\!1$ at $\theta\!=\!-\mu$ is physically expected. For the Monte-Carlo simulations presented in Sect.~\ref{Sec:MonteCarlo}, we use the $k$-range associated with rays perpendicular to the plane ($z$-direction) at $\theta\!=\!-\psi$ since this sampling direction best represents the intrinsic structural width of the spiral arm, rather than the compressed post-shock layer. It captures the effective spread of bound and marginally bound material around the arm’s density ridge, and thus provides a physically meaningful measure of the morphological arm thickness relevant for radiative transfer and dust formation modelling.

\section{Phase lag between light curve and radial velocity for an eclipsing binary system}
\begin{figure}[htp]
	\centering
	\includegraphics[width = \columnwidth]{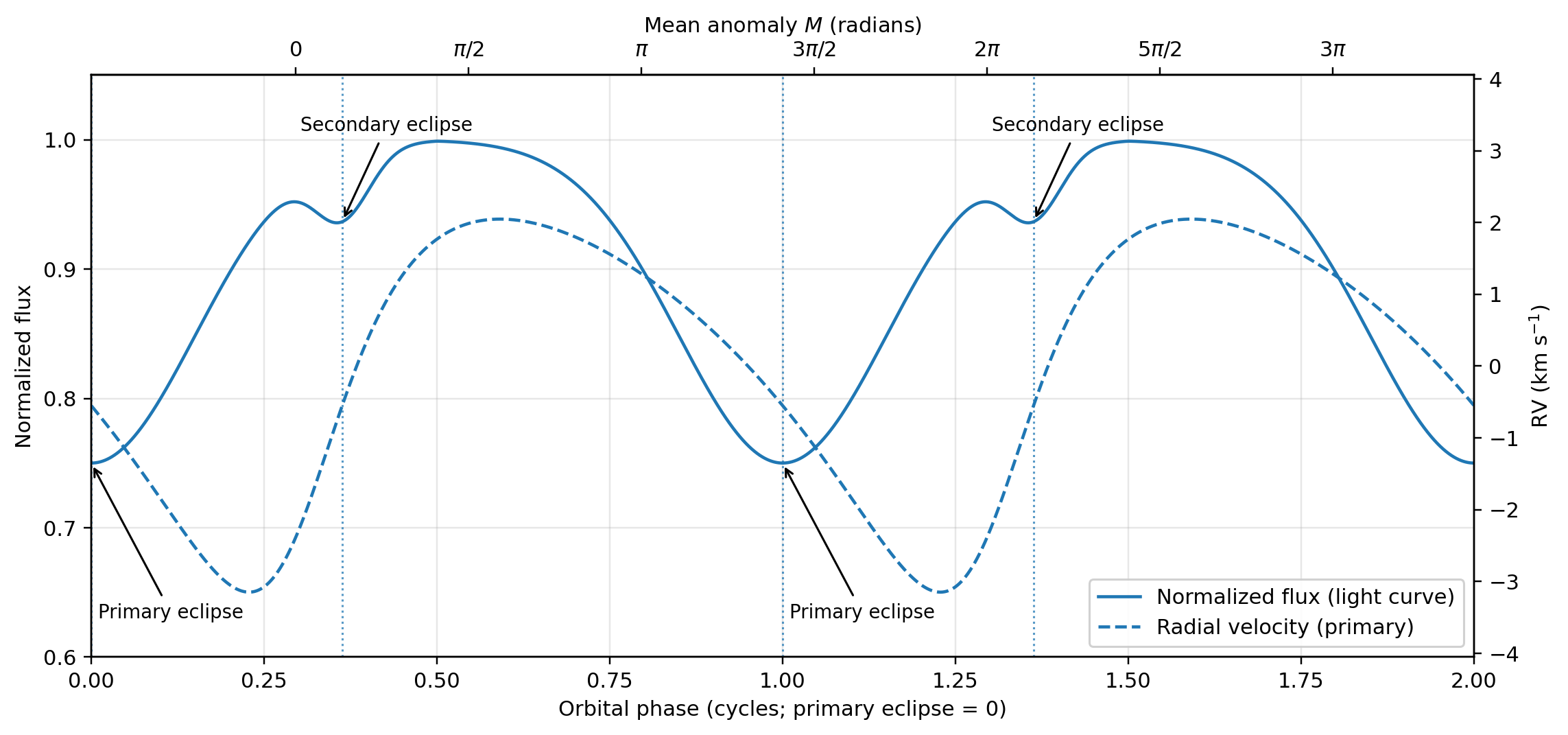}
\caption{Relation between the orbital light curve (solid) and the radial-velocity curve (dashed) for an eclipsing binary with \(e = 0.3\), \(\omega = 225^\circ\), and \(i = 60^\circ\). The bottom axis shows the normalized orbital phase, while the top axis indicates the corresponding mean anomaly \(M\). Dotted vertical lines mark the epochs of primary and secondary eclipses. For a circular, edge-on orbit, the primary's RV minimum is a quarter of an orbital cycle  (\(\pi/2\)) later in phase than the  light curve minimum (primary eclipse); for eccentric orbits, this offset depends on both \(e\) and \(\omega\).}
\label{Fig:schematic_lag}
\end{figure}
	
	\section{Online material}
	
	\paragraph{Movie S1 — Eccentric companion shaping a dusty spiral.}
	
	\begin{figure}[htp]
		\centering
		 \captionsetup{type=figure,name={Video}} 
		\includegraphics[width=\linewidth]{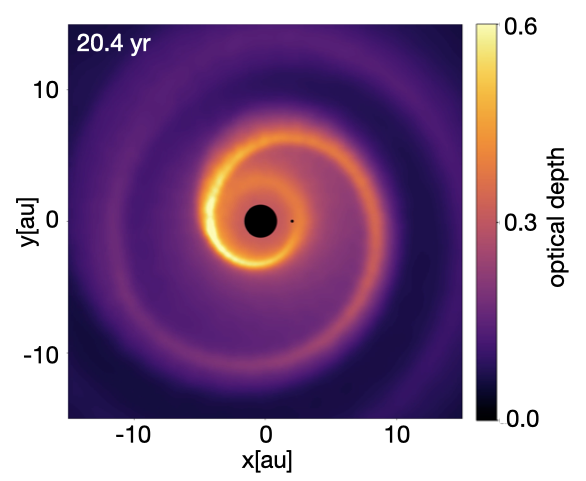}
		\caption{3D hydrodynamical simulation of an eccentric binary with a red-giant primary (\(M_1=1.5\,M_\odot\)) and a low-mass companion (\(M_2=0.125\,M_\odot\)). The orbit has semi-major axis \(a=2.3\,\mathrm{au}\) and eccentricity \(e=0.3\); the dust–condensation radius is \(R_{\rm cond}=3\,\mathrm{au}\), the orbital period is 1000 days.  The video shows the optical-depth maps in a slice through the orbital plane. The red-giant primary is at the origin \((0,0)\); the low-mass companion lies to the right in each panel. \emph{If the embedded video does not render, a copy is available on request from \texttt{leen.decin@kuleuven.be}.}}
		\label{Video:optical_depth}
	\end{figure}

\end{appendix}

\end{document}